\documentclass[aps,prx,superscriptaddress,twocolumn,amsmath]{revtex4-1}

\usepackage[utf8]{inputenc}
\usepackage{graphicx}
\usepackage{amsfonts,amssymb,amsmath}
\usepackage{siunitx}
\usepackage{mhchem}

\begin{document}

\title{Superconductivity in Sn$_{1-x}$In$_{x}$Te thin films grown by molecular beam epitaxy}

\author{Andrea Bliesener}
\thanks{These authors contributed equally to this work.}
\author{Junya Feng}
\thanks{These authors contributed equally to this work.}
\author{A. A. Taskin}
\author{Yoichi Ando}
\email[]{ando@ph2.uni-koeln.de}
\affiliation{Physics Institute II, University of Cologne, Z\"ulpicher Str. 77, 50937 K\"oln, Germany}

\begin{abstract}
The superconductor Sn$_{1-x}$In$_{x}$Te is derived from the topological crystalline insulator SnTe and is a candidate topological superconductor. So far, high-quality thin films of this material have not been available, even though such samples  would be useful for addressing the nature of its superconductivity. Here we report the successful molecular beam epitaxy growth of superconducting Sn$_{1-x}$In$_{x}$Te films by using Bi$_2$Te$_3$ as a buffer layer. The data obtained from tunnel junctions made on such films show the appearance of two superconducting gaps, which points to the coexistence of bulk and surface superconductivity. Given the spin-momentum locking of the surface states, the surface superconductivity is expected to be topological with an effective $p$-wave character. Since the topological surface states of SnTe consist of four Dirac cones, this platform offers an interesting playground for studying topological surface superconductivity with additional degrees of freedom.

\end{abstract}
\maketitle

\section{Introduction}

Recently, topological superconductors (TSCs) \cite{Qi2011,Sato2017} have aroused significant interest as an important class of topological materials. Due to the bulk-edge correspondence \cite{Ando2013}, a gapped topological bulk is always accompanied by a gapless boundary, which may harbor emergent Majorana states in the case of TSCs \cite{Sato2017}. The Majorana states are peculiar in that their creation and annihilation operators are identical, which means in the quantum field theory that particles are their own antiparticles \cite{Wilczek2009}. When localized, the Majorana states become Majorana zero modes and obey non-Abelian statistics, allowing for topological quantum computation \cite{Alicea2012}. This is one of the reasons why TSCs attract so much attention. Nevertheless, concrete materials that have been established to be TSCs are rare \cite{Sato2017}, and the search for TSCs is forming a major trend in the materials-oriented research in current condensed matter physics.

Among the candidate TSC materials, \ce{Sn$_{1-x}$In$_{x}$Te} \cite{Erickson2009}, which is derived from the topological crystalline insulator SnTe \cite{Tanaka2012}, is peculiar because it has been proposed that its bulk superconducting (SC) state can be both an odd-parity $p$-wave state or an even-parity $s$-wave state, depending on the disorder \cite{Novak2013}; namely, it was speculated that only when the sample is clean enough, the bulk presents the $p$-wave SC states which gives rise to a pronounced zero-bias conductance peak (ZBCP) in the point-contact spectroscopy \cite{Sasaki2012}, while the pairing loses the unconventional nature when the sample is disordered \cite{Novak2013, Maeda2017}. Nonetheless, even in the latter case, the topological surface states stemming from the topological-crystalline-insulator nature of SnTe remain intact after In doping to induce superconductivity \cite{Sato2013}; through the proximity effect from the bulk superconductivity, these surface states will obtain a superconducting gap \cite{Hosur2011}, presenting an effective $p$-wave two-dimensional (2D) superconductivity on the surface in a similar manner to that in the proposal by Fu and Kane \cite{Fu2008}. 

Therefore, \ce{Sn$_{1-x}$In$_{x}$Te} is a very interesting system in the context of TSCs and its detailed studies involving various types of devices (such as Josephson junctions, tunnel junctions, superconducting quantum interference device  etc.) would be useful. In this regard, \ce{Sn$_{1-x}$In$_{x}$Te} has a rock-salt structure and is difficult to be exfoliated into thin flakes, making high-quality thin film samples to have particular importance for device fabrications. To the best of our knowledge, it has not been possible to grow superconducting epitaxial \ce{Sn$_{1-x}$In$_{x}$Te} thin films using the molecular beam epitaxy (MBE) method. In this Rapid Communication, we report the successful growth of such thin films, which became possible by the use of a suitable buffer layer and employing {\it in-situ} annealing. Furthermore, we report the results of tunnel junction devices. The tunneling spectroscopy data present two gaps without any ZBCP, pointing to the scenario that the bulk is a conventional superconductor which proximitizes the topological surface states to also become superconducting.

\section{Experiment}
\subsection{Thin film growth}
The \ce{Sn$_{1-x}$In$_{x}$Te} films were grown on epi-ready sapphire (0001) substrates using the MBE method in an ultrahigh vacuum chamber with the background pressure better than 1 $\times$ 10$^{-7}$ Pa. As the first step, 15-nm-thick \ce{Bi2Te3} was grown as a buffer layer while ramping the sample temperature from \SI{210}{\celsius} to \SI{260}{\celsius}. In the second step, a layer of \ce{Sn$_{1-x}$In$_{x}$Te} was grown on top of \ce{Bi2Te3} at \SI{260}{\celsius}, and the grown film was annealed at \SI{320}{\celsius} {\it in-situ} in the presence of Te vapor; as we explain later, this {\it in-situ} annealing in the MBE chamber immediately after the growth was necessary for obtaining superconducting films. 

\begin{figure}
\includegraphics[width=8.5cm]{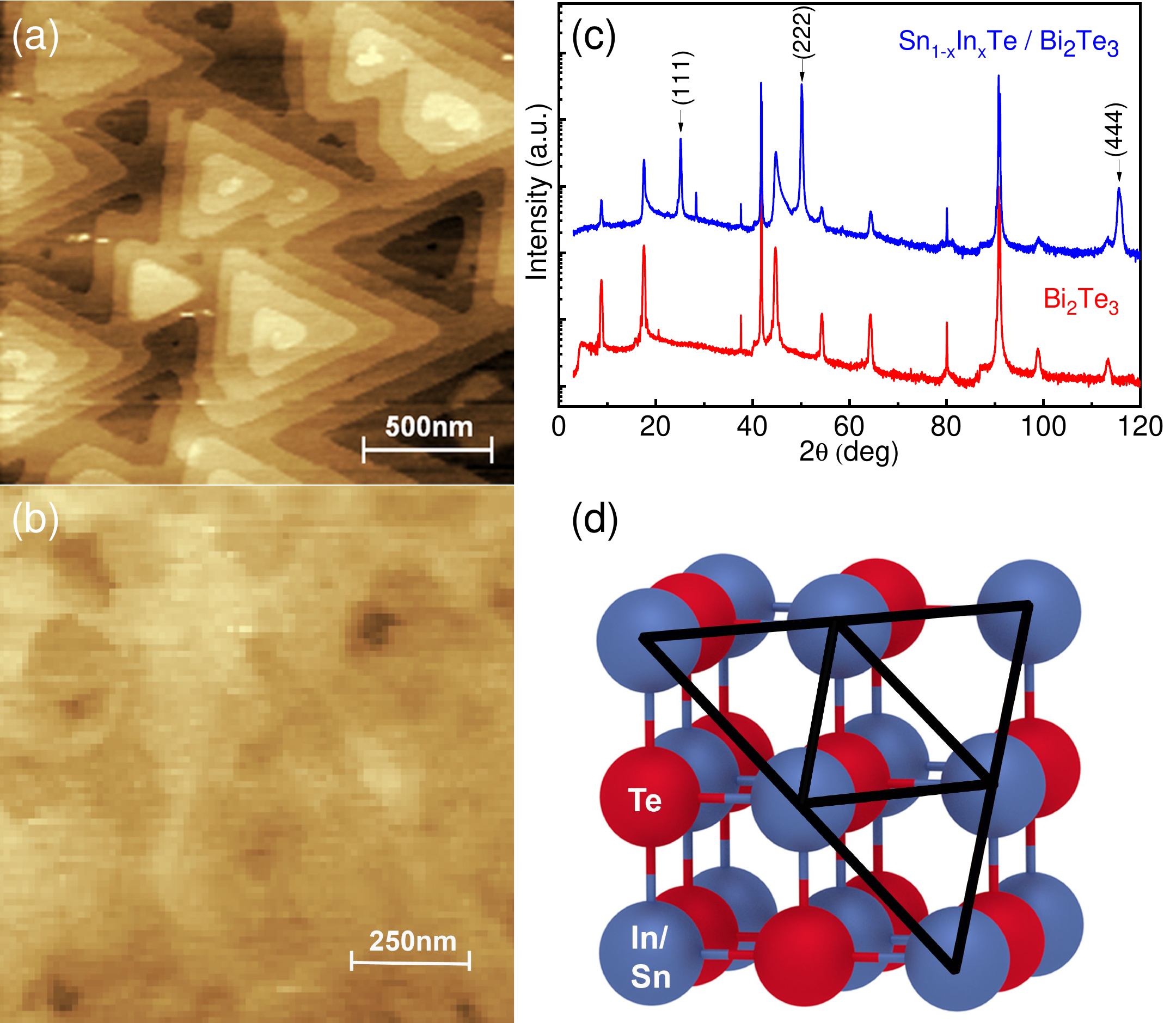}
\caption{Structural characterization of the grown film. (a) Atomic force microscopy (AFM) image of a \ce{Bi2Te3} film grown as a buffer layer. (b) AFM image of the top surface of a \ce{Sn$_{1-x}$In$_{x}$Te}/\ce{Bi2Te3} heterostructure. (c) XRD pattern of \ce{Bi2Te3} buffer layer (red) and the \ce{Sn$_{1-x}$In$_{x}$Te}/\ce{Bi2Te3} heterostructure (blue); the arrows mark the (LLL) peaks of \ce{Sn$_{1-x}$In$_{x}$Te}. (d) Rocksalt crystal structure of \ce{Sn$_{1-x}$In$_{x}$Te} with the (111) plane consisting of Sn atoms depicted by black triangles.}
\label{fig:structure}
\end{figure}

The growth rate of the \ce{Bi2Te3} buffer layer was approximately 1.25 nm/min, while it was 0.5 nm/min for the \ce{Sn$_{1-x}$In$_{x}$Te} layer.
The use of \ce{Bi2Te3} as the buffer layer for SnTe-MBE was invented by Taskin {\it et al.} \cite{Taskin2014} based on the  recognition that the (111) plane of \ce{SnTe} fits the final hexagonal \ce{Te}-plane of the \ce{Bi2Te3} quintuple layer [see Fig. \ref{fig:structure}(d)]. 

The composition of the grown and annealed films, in particular the concentration of the In dopant $x$, was analyzed by using the energy-dispersive x-ray spectroscopy (EDX) in a scanning electron microscope. Note that the coexistence of the \ce{Bi2Te3} buffer layer made the determination of the Sn/Te ratio in the \ce{Sn$_{1-x}$In$_{x}$Te} layer through the EDX analysis impossible.
The crystal structure of the grown films as well as the epitaxial nature of the growth were confirmed by using x-ray diffraction (XRD). 

\subsection{Device fabrication}

After the growth of Sn$_{1-x}$In$_{x}$Te/Bi$_{2}$Te$_{3}$ films, photolithography and Ar dry etching were employed to define strips with $\sim$20 $\mu$m width. Then, over-exposed PMMA with small windows written by electron-beam lithography (EBL) was used to cover a part of the strip and define the tunnel-junction area. After using O$_{2}$ reactive-ion etching and weak Ar etching to clean the surface of the area exposed for tunnel junctions, a thin (0.27 nm) Al$_{2}$O$_{3}$ layer was deposited as the tunnel barrier with the atomic layer deposition technique using Ultratec Savannah S200. Finally, the tunnel-junction electrodes were defined with another step of EBL, with sputtering and lift-off of a Pt/Au layer (2 nm/120 nm). Throughout the device fabrication, the samples were never heated to above \SI{120}{\celsius}, so that the properties of the films were not affected by the process.

\subsection{Measurements}
Transport measurements were performed either in a dry dilution refrigerator (Oxford Instruments Triton 200) with a base temperature of 8 mK or in a helium-3 insert (Oxford Instruments Heliox) in a 14-T superconducting magnet. The ac lock-in technique was used for measuring the resistance and the Hall resistance. In the tunneling spectroscopy experiments, the bias voltage was generated by a Keithley 2450 source meter together with a voltage divider. The tunneling current was measured by a Keithley 2182A nanovoltmeter after amplified by a FEMTO DLCPA-200 current amplifier.

\section{Results}

\subsection{Sample characterizations}

\begin{figure}
\includegraphics[width=6.5cm]{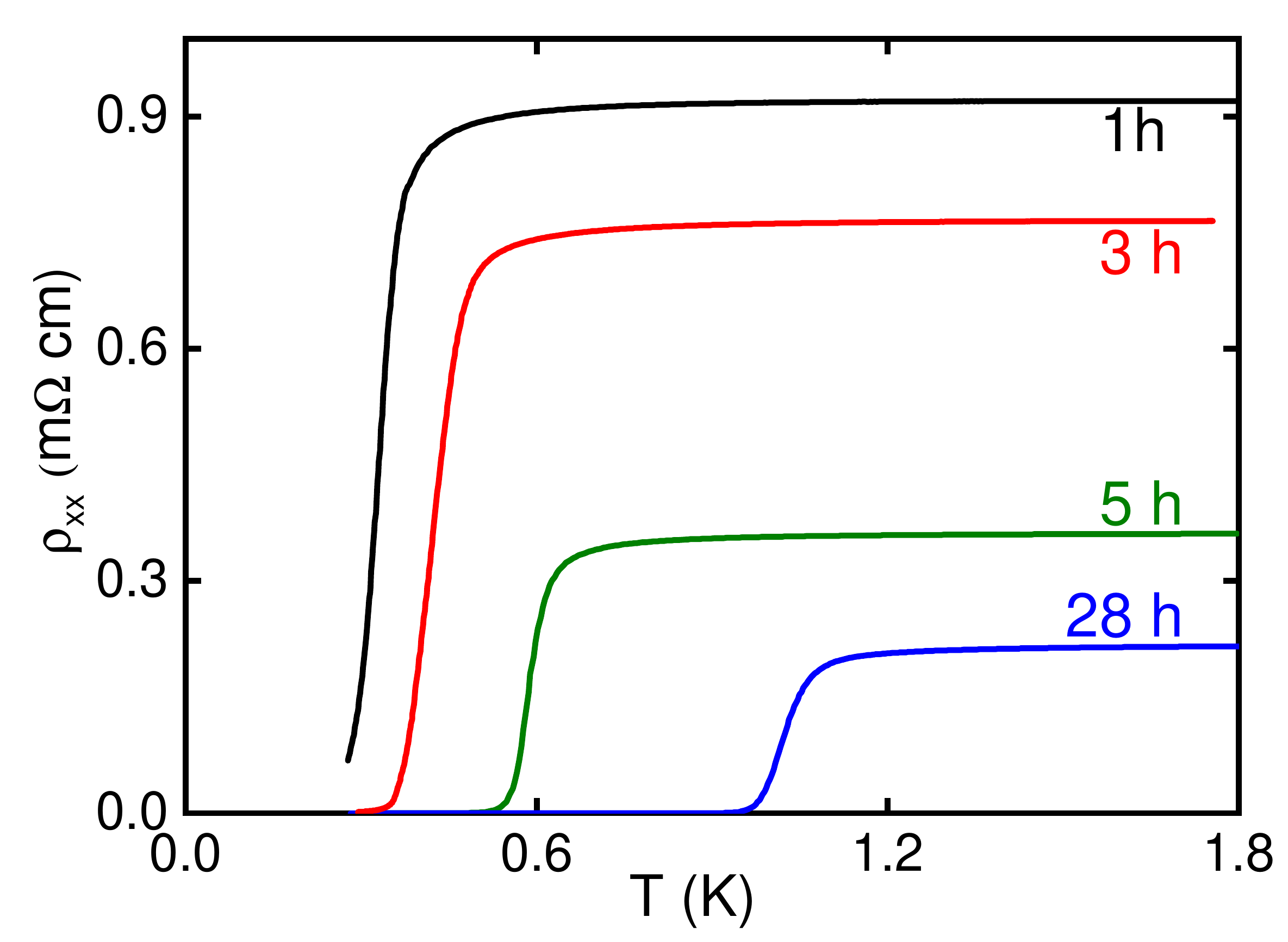}
\caption{Resistivity vs temperature near $T_c$ measured on a series of \ce{Sn$_{1-x}$In$_{x}$Te}/\ce{Bi2Te3} films annealed at \SI{320}{\celsius} for different annealing times indicated on the curves; all had a Sn$_{1-x}$In$_{x}$Te layer grown for 1 h. The $x$ values were 0.13, 0.13, 0.12, and 0.16 for the samples with the annealing time of 1, 3, 5, and 28 h, respectively. The Hall carrier density $n_{\rm H}$ was measured on these films, but it is a convolution of the hole carriers in the 30-nm-thick \ce{Sn$_{1-x}$In$_{x}$Te} layer and the electron carriers in the 15-nm-thick \ce{Bi2Te3} layer (about 4 $\times$ 10$^{19}$ cm$^{-3}$ \cite{Taskin2014}), and hence it overestimates the actual hole carrier density $p$ in \ce{Sn$_{1-x}$In$_{x}$Te} \cite{noteHall}; nevertheless, we observed $n_{\rm H}$ of 1.4, 1.2, 2.2, and 5.5 $\times$ 10$^{21}$ cm$^{-3}$ for the 1-h, 3-h, 5-h, and 28-h annealed sample, respectively, which suggests that the actual $p$ tends to increase with annealing time.
}
\label{fig:various}
\end{figure}

\begin{figure*}
\includegraphics[width=16.5cm]{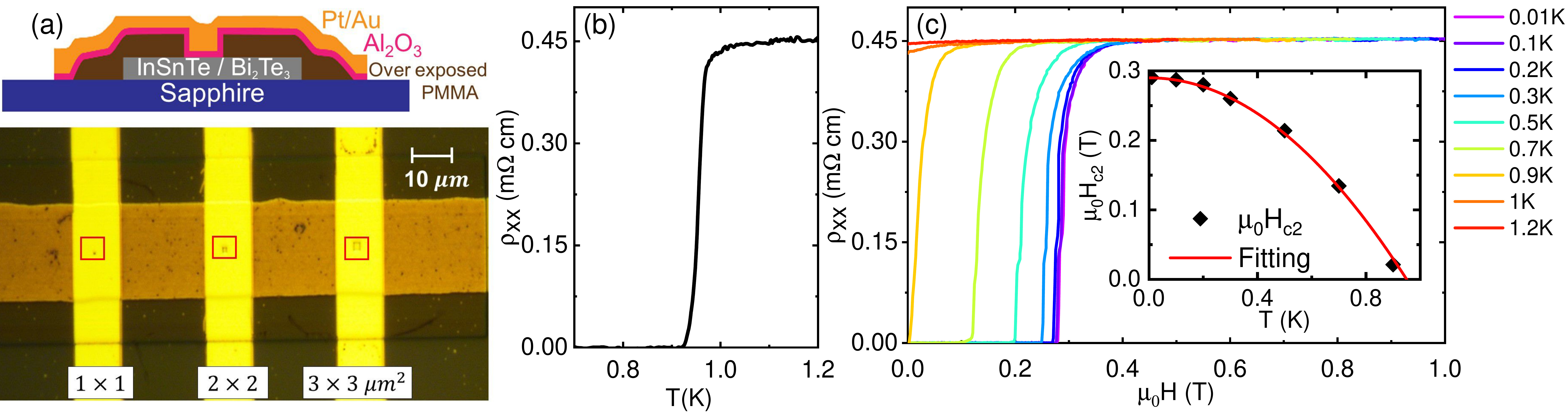}
\caption{(a) A side-view sketch of the construction of the tunnel-junction device (top) and the optical-microscope photograph of the device (bottom) with the location of the tunnel junctions marked by red squares. (b) Resistive transition of the \ce{Sn$_{0.86}$In$_{0.14}$Te}/\ce{Bi2Te3} strip on which the junctions are made; this 170-nm-thick film was annealed for 5 h at \SI{320}{\celsius}.  (c) Magnetic-field dependencies of the resistivity at various temperatures; inset shows the temperature dependence of $H_{c2}$ with the theoretical WHH curve.
}
\label{fig:rtrb2}
\end{figure*}

Generally, the \ce{Sn$_{1-x}$In$_{x}$Te} films are not superconducting in the as-grown state. The crucial step for successfully obtaining superconducting Sn$_{1-x}$In$_{x}$Te films is the {\it in-situ} annealing in the MBE chamber immediately after the growth. In our experiments, many samples were grown and annealed under various conditions, resulting in the critical temperature $T_{c}$ (defined as the mid-point in the resistive transition) ranging from 0.3 to 2.0 K. The $T_{c}$ depends also on the In content $x$, with the highest $T_{c}$ observed for $x \simeq$ 0.12 (see Supplemental Material). The error in $x$ is within the last significant digit. Interestingly, the superconducting properties of the films change with the annealing time and temperature, which suggests that some thermal activation process is involved in turning non-superconducting films into superconducting films. 

Figure \ref{fig:various} shows the resistive transitions of a series of samples annealed at \SI{320}{\celsius} for different annealing times. The $T_{c}$ tends to become higher with longer annealing time (a plot of $T_c$ vs annealing time is shown in Supplemental Material for two series of samples). At the same time, although the hole carrier density $p$ in the \ce{Sn$_{1-x}$In$_{x}$Te} layer of our films cannot be reliably extracted from the Hall resistivity due to the coexisting \ce{Bi2Te3} layer which contributes electron carriers \cite{noteHall}, one can infer from the effective Hall carrier density $n_{\rm H}$ = $(eR_{\rm H})^{-1}$ with $R_{\rm H}$ the Hall coefficient at low fields] that $p$ tends to increase with annealing time. 

This observation suggests that the In atoms in as-grown films contribute little to the carrier doping in SnTe and that the annealing promotes the doping of hole carriers. The exact chemistry of the annealing process is beyond the scope of the present work, but we speculate that due to the low growth temperature, In atoms in as-grown films sit at some non-equilibrium positions where they are inactive; after moving to the equilibrium position upon annealing, they become able to provide hole carriers and contribute to the superconductivity. In this regard, it is useful to note that the vapor-transport growth of \ce{Sn$_{1-x}$In$_{x}$Te} single crystals in Ref. \onlinecite{Novak2013} was done at \SI{630}{\celsius}, while the present MBE growth was done at \SI{260}{\celsius}. Note also that the $T_{c}$ of our \ce{Sn$_{1-x}$In$_{x}$Te} films are generally lower than that of single crystals with the same $x$ values \cite{Novak2013, Zhong2013}, suggesting that some In atoms always remain at non-equilibrium positions.

The structural quality of the as-grown films was checked with atomic force microscopy (AFM) and XRD measurements. Figure \ref{fig:structure}(a) shows a topographic AFM image of a \ce{Bi2Te3} film grown with the same conditions as the buffer layer used for the heterostructures. Triangular terraces are clearly visible on the surface of the film, indicating high quality. In Fig.~\ref{fig:structure}(b) one can see that the top surface of the \ce{Sn$_{1-x}$In$_{x}$Te} layer is essentially flat without any holes. The XRD pattern taken on a typical \ce{Sn$_{1-x}$In$_{x}$Te}/\ce{Bi2Te3} heterostructure is shown in Fig. \ref{fig:structure}(c), which demonstrates that only the (LLL) peaks of \ce{Sn$_{1-x}$In$_{x}$Te} shows up on top of the \ce{Bi2Te3} peaks in the data, affirming an epitaxial growth. The lattice constant extracted from the XRD data was 0.6310$\pm$0.0002 nm, which is essentially the same as that of the bulk SnTe (0.632 nm).

The tunnel junctions were fabricated to examine the superconducting properties of \ce{Sn$_{1-x}$In$_{x}$Te}. Tunneling spectroscopy probes the density of states (DOS) at the surface, which allows us to infer the energy gap in the SC state. A picture of the device, which is made from a 170-nm-thick $x$ = 0.14 film and is reported in the following, is shown in Fig.~\ref{fig:rtrb2}(a). The $\rho_{xx}(T)$ data of the strip on which the tunnel junctions are made is shown in Fig. \ref{fig:rtrb2}(b). The observed mid-point $T_{c}$ of 0.96 K  is typical for the films grown in this work. The Hall measurements of this film gave $n_{\rm H} \simeq$ 5 $\times$ 10$^{20}$ cm$^{-3}$, which is expected to give a reasonably good approximation of $p$ (because the \ce{Sn$_{1-x}$In$_{x}$Te} layer was ten times thicker than the \ce{Bi2Te3} layer) and is comparable to that of superconducting \ce{Sn$_{1-x}$In$_{x}$Te} single crystals with $x \simeq$ 0.04 \cite{Novak2013}.

Figure~\ref{fig:rtrb2}(c) shows the magnetic-field dependence of $\rho_{xx}$ in this device, from which we extract the upper critical field $H_{c2}$ (defined as the mid-point in the resistive transition).  From its temperature dependence [Fig. \ref{fig:rtrb2}(c) inset], we obtain the zero-temperature-limit value, $H_{c2}^0$, of 0.28 T, which is in line with the result on single crystals with similar $T_c$'s \cite{Sasaki2012}. The $H_{c2}(T)$ behavior can be fitted well with the Werthamer-Helfand-Hohenberg (WHH) theory \cite{Werthamer1966} with $\mu_{0} H_{c2}^0=-0.693 T_{c}(dH_{c2}/dT)_{T=T_c}$, suggesting a conventional nature of the bulk superconductivity.

\subsection{Tunneling spectroscopy}

\begin{figure}
\includegraphics[width=6.5cm]{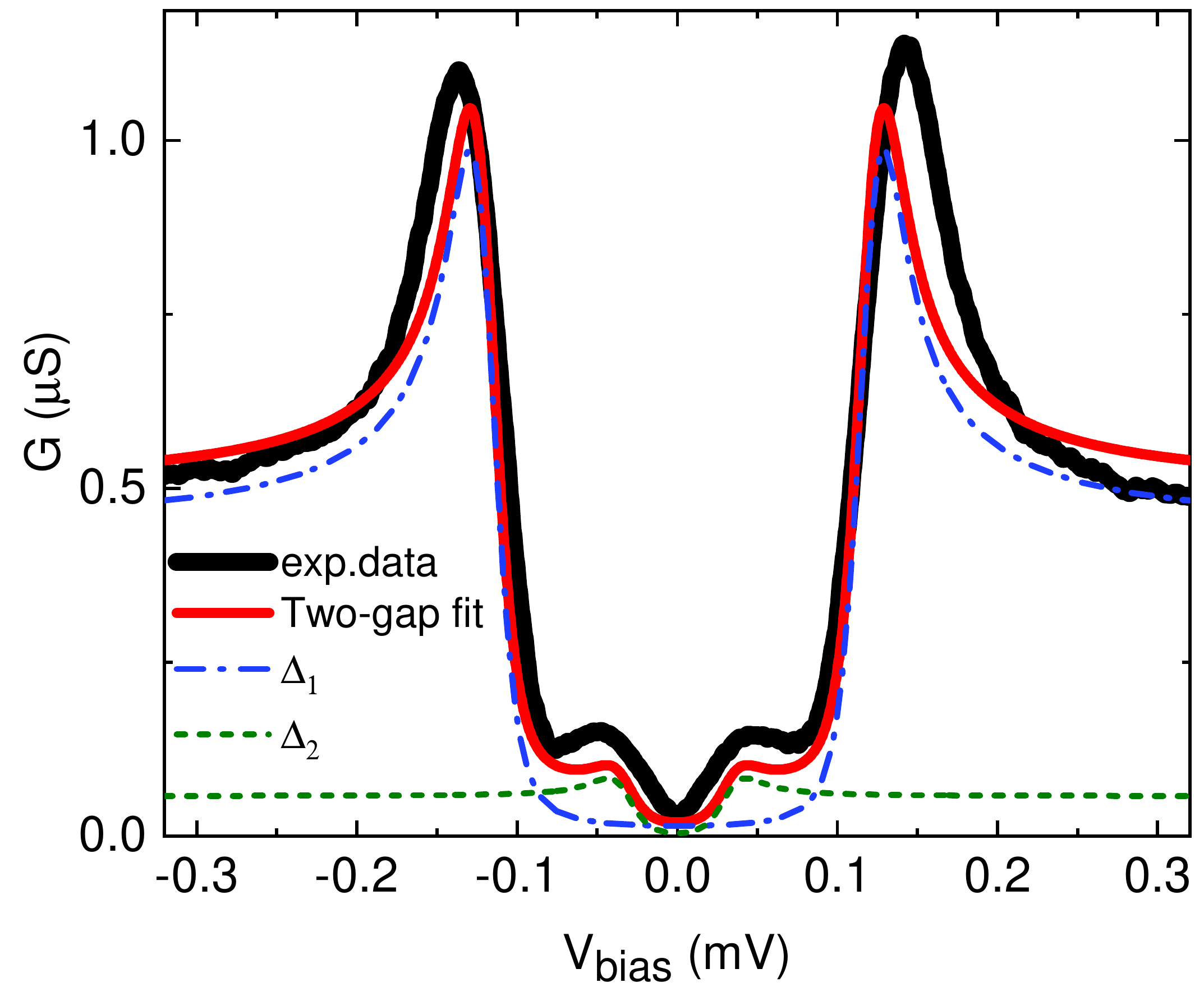}
\caption{Tunneling spectrum measured at 8 mK (black curve) and the calculated curves for two different gaps (blue and green dashed curves) as well as their superposition (red curve). The parameters used for the calculations are the following: (bulk) $\Delta_1$ = 0.1200$\pm$0.0003 meV, $\Gamma_1$ = 0.004$\pm$0.001, $G_{\rm N,1}$ = 0.447$\pm$0.004 $\mu$S; (surface) $\Delta_2$ = 0.0324$\pm$0.0032 meV, $\Gamma_2$ = 0.0012$\pm$0.0004, $G_{\rm N,2}$ = 0.058$\pm$0.004 $\mu$S.
}
\label{fig:tunneljuncfit}
\end{figure}

A tunneling spectrum measured at 8 mK in zero field is shown in Fig. \ref{fig:tunneljuncfit}. Interestingly, this spectrum exhibits two gaps. Given the fact that the parent material SnTe has topological surface states which are preserved after In doping \cite{Sato2013}, it is possible that the spectrum reflects the coexistence of bulk and surface superconductivity with different gaps.
In fact, the experimental data can be fitted with two individual BCS gaps (blue and green dashed lines) using the following Dynes formula \cite{Dynes1978},
\begin{equation}
G = \frac{\partial I}{\partial V} = G_{\rm N}\frac{\partial}{\partial V}\int_{-\infty}^{\infty}\rho(E)[f(E) - f(E-eV)]dE,
\label{eqn:BCS}
\end{equation}
where $G_{\rm N}$ is the normal-state conductance, $f(E)$ is the Fermi function, and $\rho(E)$ is the density of states in the BCS theory given by
\begin{equation}
\rho(E) = Re\left[ \frac{|E-i\Gamma|}{((E-i\Gamma)^{2} - \Delta^{2})^{1/2}}\right].
\label{eqn:BCS2}
\end{equation}
Here, $\Delta$ is the superconducting gap and $\Gamma$ is the non-dimensional broadening parameter due to a finite quasi-particle lifetime.
By considering two gaps and tuning the parameters of Eqs. (1) and (2), we obtained the red curve in Fig. \ref{fig:tunneljuncfit} which describes the observed spectrum reasonably well. The magnitudes of the gaps thus derived are $\Delta_{1}$ = 0.1200$\pm$0.0003 meV and $\Delta_{2}$ = 0.0324$\pm$0.0032 meV. It is fair to say that the calculated curve does not match the experimental data perfectly; this may partly be due to the fact that the topological surface states of SnTe on the (111) surface consist of two different kinds of Dirac cones \cite{Taskin2014, Tanaka2013}, which may lead to a total of three different SC gaps.

\begin{figure}
\includegraphics[width=8.5cm]{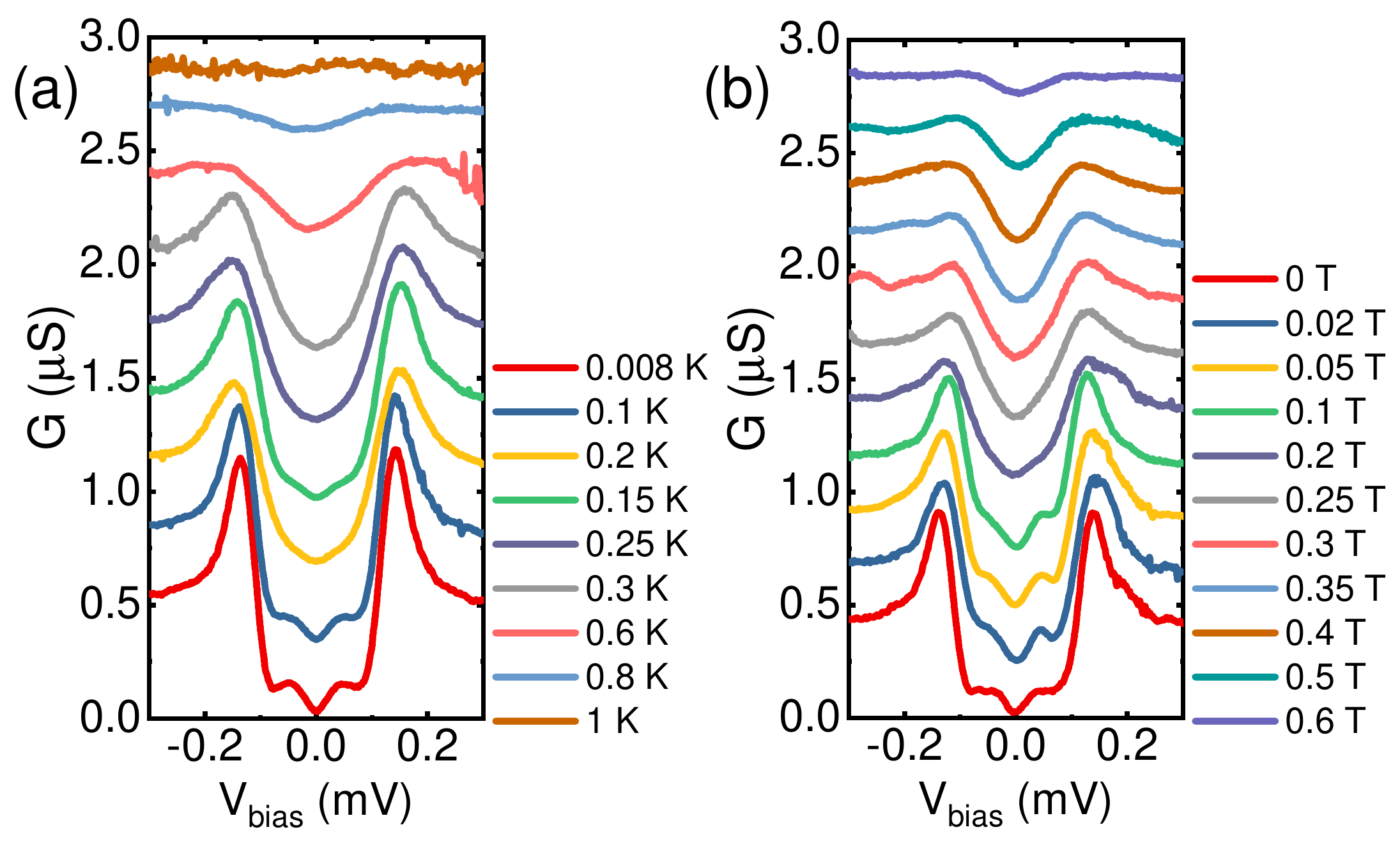}
\caption{Evolution of the tunneling spectra with (a) temperature and (b) magnetic field. The spectra are offset by \SI{0.23}{\micro\siemens} for clarity.}
\label{fig:tunneljunc}
\end{figure}

The evolution of the tunneling spectra with temperature and magnetic field is shown in Fig. \ref{fig:tunneljunc}. 
With increasing temperature from 8 mK [Fig. \ref{fig:tunneljunc}(a)], the smaller gap disappears first at 0.2 K, and then the larger gap vanishes across $T_{c}$ (= 0.95 K); in the normal state at 1.0 K, the spectrum shows no gap.
When the magnetic field is applied perpendicular to the film at 8 mK [Fig. \ref{fig:tunneljunc}(b)], the small-gap feature disappears at 0.2 T, which is below the $H_{c2}$ of 0.28 T at this temperature. Intriguingly, the larger gap remains visible above $H_{c2}$ even in 0.6 T, which suggests that the destruction of superconductivity in magnetic field occurs through the loss of phase coherence. 

\section{Discussions}

Now we discuss the implications of the observed two-gap structure in the tunneling spectra.
First, the absence of ZBCP points to the conventional $s$-wave nature of the bulk superconductivity in our \ce{Sn$_{1-x}$In$_{x}$Te} thin films. This is reasonable, because it has been discussed that the unconventional odd-parity superconductivity is realized in \ce{Sn$_{1-x}$In$_{x}$Te} only when the sample is very clean \cite{Novak2013}; in fact, the ZBCP in the point-contact spectroscopy, which is a signature of the surface Andreev boundstates and hence points to unconventional superconductivity, has been observed only in samples with the residual resistivity of $\sim$0.15 m$\Omega$cm \cite{Sasaki2012}. Given that our thin film used for the tunnel junction device shows the residual resistivity of $\sim$0.45 m$\Omega$cm, one would not expect the unconventional bulk SC state to be realized here. 

\begin{figure}
\includegraphics[width=8.5cm]{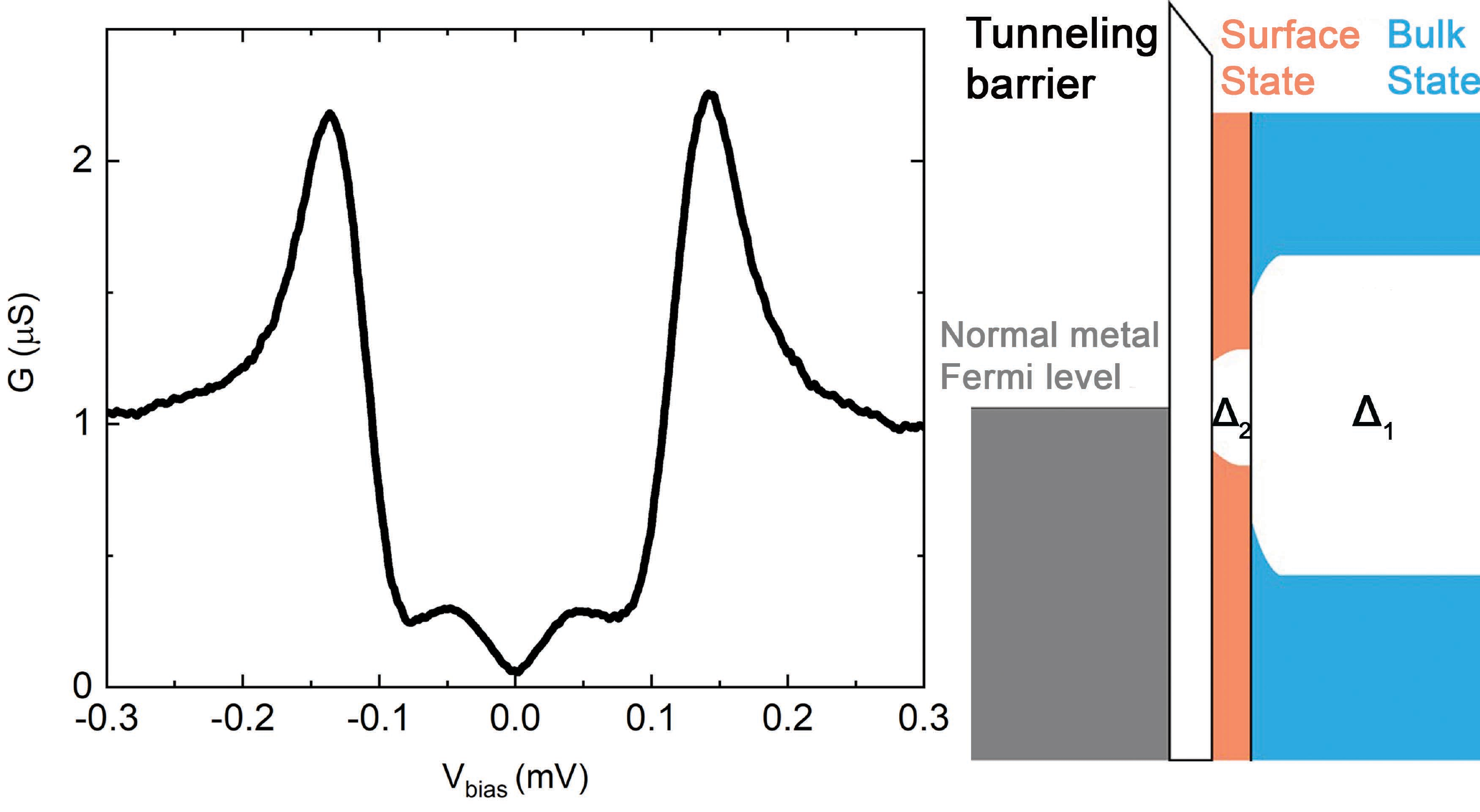}
\caption{Reproduction of the characteristic two-gap tunneling spectrum (left) and a sketch of the corresponding DOS of the metal electrode, tunneling barrier, and the two superconducting gaps (right).}
\label{fig:DOS}
\end{figure}

It is also important to notice that SnTe retains its topological surface states even after sufficient In-doping to induce superconductivity, as has been shown by angle-resolved photoemission spectroscopy experiments \cite{Sato2013}. It was already discussed for Cu$_x$Bi$_2$Se$_3$ \cite{Hosur2011} that, when the bulk states become superconducting, Cooper pairing will also occur in the surface states due to the proximity effect. In this case, the corresponding surface SC gap is naturally smaller than that of the bulk. This situation is depicted in Fig. \ref{fig:DOS}.
The advantage of this situation is that the bulk states and the surface states coexist in the same material and hence there is no interface barrier to disturb the proximity effect. It is useful to mention that a similar situation has been reported to be realized in the iron-based superconductor \ce{FeTe$_{1-x}$Se$_{x}$} \cite{Zhang2018}, for which the Majorana bound states have been claimed to be observed in the vortex core \cite{Wang2018}.

It is prudent to mention that it was rather difficult to observe the clear two-gap feature in the tunneling spectra, although we have seen it in more than one device. To be specific, we measured a total of 13 junctions, among which 3 showed a clear double gap feature and 6 showed a V-shaped gap suggestive of the two gaps; the other 4 showed the usual U-shaped gap. 
It appears either that the surface SC gap can be easily smeared due to disorder or that the wave functions of the 2D surface states are much more difficult to be accessed through a tunnel barrier. This in turn supports the surface origin of the small gap and provides evidence against the possibility that \ce{Sn$_{1-x}$In$_{x}$Te} is a bulk two-band superconductor and that the smaller gap is associated with a minor bulk band.  

From the materials perspective, it is useful to clarify the possible impact of Sn vacancies, because they are the main source of hole doping in pristine SnTe. In this regard, we have extensively studied pristine SnTe films grown on Bi$_2$Te$_3$ buffer layer \cite{Taskin2014}, and the typical hole density in pristine SnTe layers grown at \SI{320}{\celsius} (the annealing temperature of this study) was $\sim$2~$\times$~10$^{20}$ cm$^{-3}$, which is at the lower border for superconductivity \cite{Novak2013}. Since the hole density of our superconducting samples was generally larger (e.g., it was $\sim$5~$\times$~10$^{20}$ cm$^{-3}$ in the sample shown in Fig. 3), it would be reasonable to conclude that the Sn vacancies are not playing a dominant role in the occurrence of superconductivity. 

One should note that the proximity-induced surface superconductivity in our system is bound to have the effective $p$-wave character, because the four Dirac cones forming the topological surface states are characterized by spin-momentum locking and the resulting pairing is effectively spinless \cite{Fu2008}. An important difference from the case of the $Z_2$ topological insulators is that there is more than one Dirac cone; in the case of the (111) surface of SnTe, there is one Dirac cone at the center of the Brillouin zone ($\bar{\Gamma}$ point) plus three additional Dirac cones at the Brillouin zone boundary ($\bar{M}$ points) \cite{Tanaka2013}. It is an interesting future topic to elucidate whether/how the Majorana zero modes associated with such a band structure hybridize when they are created in a vortex core. Also, the effect of uniaxial strain, which will break mirror symmetry and gap out some of the surface Dirac cones, would be interesting to study in this platform.

\section{Summary}

By employing \ce{Bi2Te3} as a buffer layer, we have been able to grow superconducting \ce{Sn$_{1-x}$In$_{x}$Te} epitaxial thin films with the MBE method; however, {\it in-situ} annealing of the film for a long time is necessary for observing superconductivity, and $T_c$ is generally lower than that in single crystals. We observed that the hole carrier density inferred from the Hall resistivity tends to increase with longer annealing time, and so is $T_c$. The residual resistivity of our superconducting films is larger than the cleanest single crystals, which probably causes the bulk SC state to have conventional $s$-wave nature. Nevertheless, our tunneling spectroscopy data strongly suggests that the topological surface states consisting of four Dirac cones are proximitized and obtain a SC gap, which is bound to result in a topological surface superconductivity due to the spin-momentum locking of the surface Dirac cones. It will be important to elucidate the Majorana physics in such a situation with additional valley degrees of freedom.

\acknowledgements{We thank Fan Yang for his support in the experiments at the initial stage. This project has received funding from the European Research Council (ERC) under the European Union's Horizon 2020 research and innovation program (Grant Agreement No. 741121) and was also funded by the Deutsche Forschungsgemeinschaft (DFG, German Research Foundation) under CRC 1238 - 277146847 (Subprojects A04 and B01) as well as under Germany's Excellence Strategy - Cluster of Excellence Matter and Light for Quantum Computing (ML4Q) EXC 2004/1 - 390534769.

\bibliography{insntepapers}

\begin{thebibliography}{21}
\bibitem [{not()}]{noteHall}%
  \BibitemOpen
  \href@noop {} {}\bibinfo {note} {When a heterostructure gives rise to
  parallel channels of electron and hole carriers, the Hall effects from the
  two channels cancel each other and result in a small effective Hall
  coefficient, which corresponds to an unrealistically large effective carrier
  density.}\BibitemShut {Stop}%
\end{thebibliography}

\end{document}


\title{Supplemental Material for\\ Superconductivity in Sn$_{1-x}$In$_{x}$Te thin films grown by molecular beam epitaxy}

\author{Andrea Bliesener}
\thanks{These authors contributed equally to this work}
\author{Junya Feng}
\thanks{These authors contributed equally to this work}
\author{A. A. Taskin}
\author{Yoichi Ando}
\email[]{ando@ph2.uni-koeln.de}
\affiliation{Physics Institute II, University of Cologne, Z\"ulpicher Str. 77, 50937 K\"oln, Germany}

\maketitle

\begin{figure} [b]
\includegraphics[width=9.5cm]{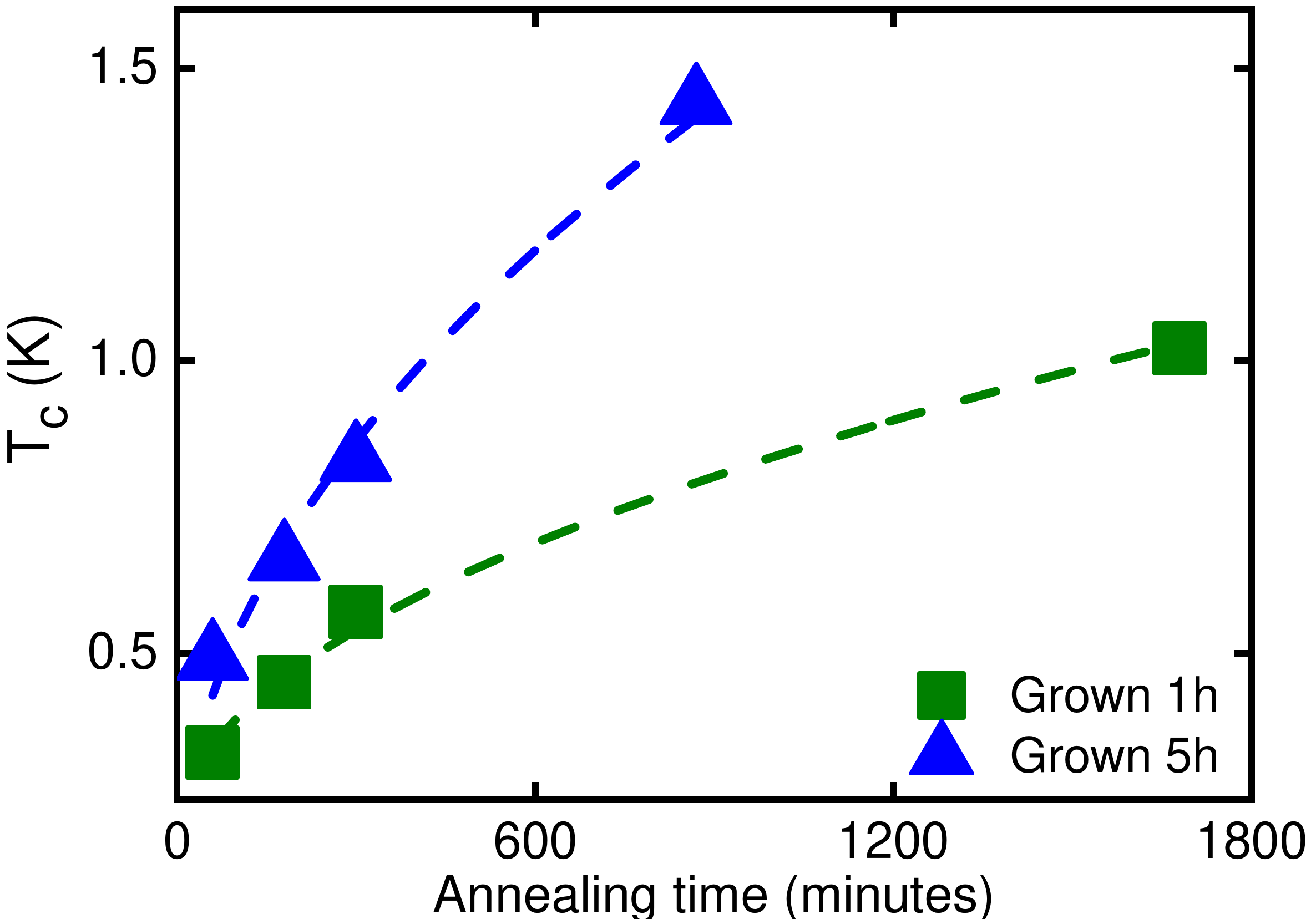}
\caption{Plots of $T_{c}$ vs annealing time for samples grown for 1 h (green squares) and 5 h (blue triangles). The samples were annealed {\it in-situ} at $320^\circ$C for various durations of time.}
\label{fig:fig1}
\end{figure}

Figure \ref{fig:fig1} shows the dependence of the critical temperature $T_{c}$ on the annealing time for two series of samples. The samples of the 1-h series were all grown at $260^\circ$C for 1 h, and then annealed {\it in-situ} at $320^\circ$C for different durations of time. The 5-h series were grown for 5 hours. The latter series shows higher $T_c$'s compared to the former. 

\begin{figure} [b]
\includegraphics[width=9.5cm]{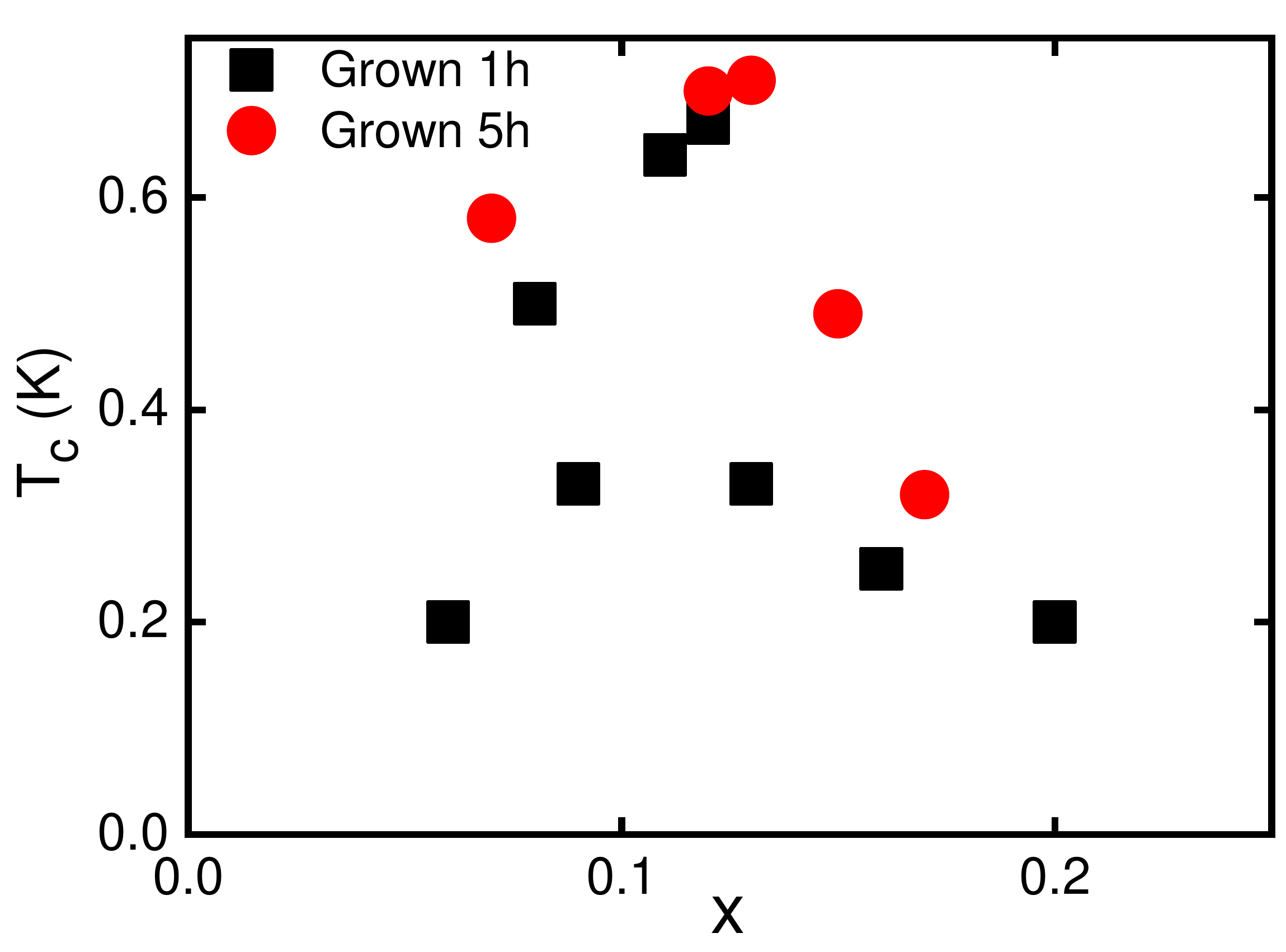}
\caption{Plots of $T_{c}$ vs In content $x$ for samples grown for 1 h (black squares) and 5 h (red circles). All samples were annealed {\it in-situ} at $320^\circ$C for 1 h.}
\label{fig:fig2}
\end{figure}

The dependence of $T_{c}$ on the In content $x$ is shown in Fig. \ref{fig:fig2} for two series of samples. The films were grown for either 1 h or 5 h, and then {\it in-situ} annealed for 1 h at $320^\circ$C. The data shows a dome-like shape with a $T_{c}$ maximum located at $x$ = 0.11 -- 0.12.

\begin{singlespace}
\begin{table} [t]
\centering

 \begin{tabular}{|c|c|c|c|c|c|c|c|c|} 
 \hline  
 & Growth   & Growth &  Annealing  &  Annealing &  $x$ &  thickness&  $T_{c}$ (K) &  $n_{\rm H}$ (cm$^{-3}$) \\ 
  &  temp. ($^\circ$C) & time (min) &   temp. ($^\circ$C) & time (min) &  &  (nm)&   &   \\ 
 

\hline

 1 & 260 & 60 & 320 & 60 & 0.13 & 55 & 0.33 & 1.4 $\times$ 10$^{21}$ \\ 

 2 & 260 & 60 & 320 & 180 & 0.13 & 55 & 0.43 & 1.2 $\times$ 10$^{21}$ \\ 

 3 & 260 & 60 & 320 & 300 & 0.12 & 45 & 0.59 & 2.2 $\times$ 10$^{21}$ \\ 

 4 & 260 & 60 & 320 & 1680 & 0.16 & 37 & 1.02 & 5.5 $\times$ 10$^{21}$ \\ 

 5 & 260 & 300 & 320 & 300 & 0.14 & 170 & 0.96 & 5.0 $\times$ 10$^{20}$ \\ 

 \hline  

\end{tabular}
\caption{Summary of the growth parameters and basic characterizations of the samples reported in the main text. Samples 1 -- 4 are shown in Fig. 2 of the main text, and Sample 5 is the film used for making the device shown in Fig. 3.}
\label{table:samples}
\end{table}
\end{singlespace}
